\documentclass[prl,showpacs,showkeys,nofootinbib,twocolumn,floatfix,superscriptaddress]{revtex4}

\usepackage{graphicx}  %
\usepackage{amsmath}
\usepackage{bm}  %
\usepackage{ulem}

\newcommand{\bea}{\begin{eqnarray}}
\newcommand{\eea}{\end{eqnarray}}
\newcommand{\beq}{\begin{equation}}
\newcommand{\eeq}{\end{equation}}
\newcommand{\bqa}{\begin{eqnarray}}
\newcommand{\eqa}{\end{eqnarray}}

\def\mqo2{{\!\!\!}}

\begin{document}

\title{
Universal Relations for Identical Bosons \\ 
from 3-Body Physics}

\author{Eric Braaten}\email{braaten@mps.ohio-state.edu}
\affiliation{Department of Physics,
         The Ohio State University, Columbus, OH\ 43210, USA\\}

\author{Daekyoung Kang}\email{kang@mps.ohio-state.edu}
\affiliation{Department of Physics,
         The Ohio State University, Columbus, OH\ 43210, USA\\}

\author{Lucas Platter}
\email{platter@chalmers.se}
\affiliation{Institute for Nuclear Theory,
University of Washington, Seattle WA\ 98195 USA\\}
\affiliation{Fundamental Physics, Chalmers University of Technology,
         41296 G\"oteborg, Sweden }
\date{\today}

\begin{abstract}
Systems consisting of identical bosons with a large scattering length 
satisfy universal relations determined by 2-body physics 
that are similar to those for fermions with two spin states.  
They require the momentum distribution 
to have a large-momentum $1/k^4$ tail and the radio-frequency
transition rate to have a high-frequency $1/\omega^{3/2}$ tail,
both of which are proportional to the 2-body contact.
Identical bosons also satisfy additional universal relations 
that are determined by 3-body physics and involve the 3-body contact,
which measures the probability of 3 particles being very close together.
The coefficients 
of the 3-body contact in the $1/k^5$ tail of the momentum 
distribution and in the $1/\omega^{2}$ tail of the radio-frequency
transition rate are log-periodic functions of 
$k$ and $\omega$ that depend on the Efimov parameter.
\end{abstract}

\smallskip
\pacs{31.15.-p,34.50.-s, 67.85.Lm,03.75.Nt,03.75.Ss}
\keywords{
Bose gases, 
scattering of atoms and molecules, operator product expansion. }
\maketitle

Strongly interacting systems 
present a challenging problem in theoretical physics.
Some of the simplest of such systems consist of particles with 
short-range interactions and large scattering lengths.
They arise in almost all branches of physics, including atomic,
condensed matter, high energy, and nuclear physics.
Ultracold trapped atoms with large scattering lengths are 
particularly pristine examples of such systems.  
In addition to the exquisite probes that are available in 
atomic physics, the ability to control the scattering length 
by using Feshbach resonances makes this dimension of the system
accessible experimentally.

The simplest many-body systems of particles that interact through 
a large scattering length consist of fermions with two spin states.
In the past decade, such systems have been the subject of intensive
investigations, both theoretical and experimental, 
using ultracold trapped atoms \cite{GPS:2010}.
A powerful tool for studying these systems is universal relations 
that are determined by 2-body physics but hold for any state 
of the system \cite{Braaten:2010if}.  
These relations involve the {\it contact},
an extensive property of the system that measures the probability 
for a pair of particles in the two spin states to be very close together.
The first such relations were derived by Shina Tan \cite{Tan123}.
Universal relations were subsequently derived for 
radio-frequency spectroscopy \cite{UR:rfsum,SR:0910,Braaten:2010dv},
photoassociation \cite{UR:photo},
structure factors \cite{UR:structure},
and correlation functions related to the viscocity \cite{UR:viscosity}. 
An exciting recent development is the experimental confirmation 
of some of the universal relations \cite{UR:exp}.

In this letter, we present universal relations 
for identical bosons with large scattering length.
Like those for fermions with two spin states,
these new universal relations involve the {\it 2-body contact}.
They also involve the {\it 3-body contact}, 
an extensive property of the system that measures the probability
for triples of identical bosons to be very close together.

For identical bosons in the zero-range limit, there are two
interaction parameters: the large scattering length $a$ and an Efimov
parameter $\kappa_*$ that is defined below \cite{Braaten:2004rn}.  
Observables can depend only log-periodically on
$\kappa_*$ with discrete scaling factor $e^{\pi/s_0} \approx 22.7$,
where $s_0 \approx 1.00624$ is the solution to a transcendental equation.  
In the unitary limit $a \to \pm \infty$, there are
infinitely many Efimov trimers with an accumulation point at the
3-atom threshold.  The parameter $\kappa_*$ can be defined by
the trimer spectrum near the threshold
in the unitary limit:
$-(e^{-2 \pi/s_0})^n \hbar^2 \kappa_*^2/m$, 
where $n$ is an integer.

The 2-body contact $C_2$ and the 3-body contact $C_3$ for a 
state with energy $E$ can be defined operationally 
in terms of derivatives of the energy: 
\begin{subequations}
\begin{eqnarray}
\left( a \frac{\partial E}{ \partial a} \right)_{\!\! \kappa_*} &=& 
\frac{\hbar^2}{8\pi m a} ~ C_2,
\label{E-C2}
\\
\left( \kappa_* \frac{\partial E}{\partial \kappa_*} \right)_{\!\! a} &=& 
- \frac{2 \hbar^2}{m} ~ C_3.
\label{E-C3}
\end{eqnarray}
\label{E-C2,3}
\end{subequations}
For a many-body state at non-zero temperature, the derivatives
should be evaluated at fixed entropy.  The normalization of $C_2$ in
Eq.~(\ref{E-C2}) has been chosen so that the tail of the momentum
distribution at large wavenumber $k$ (given below in Eq.~(\ref{n-k}))
is $C_2/k^4$.  The coefficient on the right side of Eq.~(\ref{E-C2})
then differs by a factor of 1/2 from that for fermions with two spin
states.  The normalization of $C_3$ in Eq.~(\ref{E-C3}) has been
chosen so that the 3-body contacts for the Efimov trimers in the unitary
limit are $(e^{-2 \pi/s_0})^n \kappa_*^2$.  
Werner and Castin have expressed the derivative of $E$  in
Eq.~(\ref{E-C3}) in terms of the 3-body Schr\"odinger wavefunction at
small hyperradius \cite{WC:1001}.

The importance of the contacts $C_2$ and $C_3$ is that there are other
properties of the system that depend linearly on these
quantities with universal coefficients that are determined by few-body physics.
One of these universal relations gives the 
{\it tail of the momentum distribution} at large
wavevector $\bm{k}$.  We normalize the distribution $n(k)$ so that 
$\int d^3k~n(k)/(2 \pi)^3$ is the total number of atoms.
We will show below that the tail of $n(k)$ can be expressed as
\begin{equation}
n(k) \longrightarrow \frac{1}{k^4} C_2
+ \frac{F(k)}{k^5} C_3,
\label{n-k}
\end{equation}
where $F(k)$ is a universal log-periodic function:
\begin{equation}
F(k)= A~\sin[2 s_0 \ln(k/\kappa_*)+ 2 \phi] .
\label{Fmom}
\end{equation}
The numerical constants are $A = 89.2626$ 
and $\phi = -0.669064$. 
The dependence on the state enters only through the 
contacts $C_2$ and $C_3$.  
Their coefficients can be determined by matching the tail 
of the momentum distribution for any convenient state. 
Werner and Castin have calculated $C_2$ and $F(k)C_3$
for an Efimov trimer in the unitary limit
$a \to \pm \infty$ \cite{WC:1001}. 
The 2-body contact for the trimer
with binding momentum $\kappa_*$ is $C_2 = 53.0972~\kappa_*$. 
By our definition in Eq.~(\ref{E-C3}), the 3-body contact is 
$C_3 = \kappa_*^2$.  We determined $A$ and $\phi$
by matching the precise results of Ref.~\cite{WC:1001}. 
A connection between the $1/k^5$ tail in Eq.~(\ref{n-k})
and the derivative of the energy in Eq.~(\ref{E-C3})
was conjectured in Ref.~\cite{WC:1001}.

Another universal relation for identical bosons is an 
{\it Energy Relation} that expresses the energy of a system
in terms of $n(k)$, $C_2$, and $C_3$.  The total energy $E$ is the
sum of the kinetic energy $T$, the interaction energy $U$, and the
energy $V$ associated with an external trapping potential. 
The kinetic energy can be expressed as an integral 
over $n(k)$. Because of the $k^{-4}$ tail in Eq.~(\ref{n-k}),
the integral is linearly ultraviolet divergent.   
If the $k^{-4}$ tail were subtracted, the integral would be 
ultraviolet finite, but it would still depend log-periodically 
on the ultraviolet momentum cutoff because of the $k^{-5}$ tail. 
The linear and log-periodic dependence on the ultraviolet cutoff 
are both cancelled by $U$.   
The Energy Relation, which is derived below, expresses the sum $T+U$ 
in a form that is insensitive to the ultraviolet cutoff:
\begin{eqnarray}
T + U =
\int \!\! \frac{d^3k}{(2 \pi)^3}~\frac{\hbar^2 k^2}{2 m}
\left[ n(k) - \frac{C_2}{k^4}  \right.  \hspace{1.5cm}
\nonumber
\\ 
\left.  - \theta(k - k_0) \frac{F(k) C_3}{k^5} \right]
\nonumber
\\ 
+ \frac{\hbar^2}{8 \pi m a} C_2
+ \frac{[F(e^{\pi/4 s_0} k_0) + f_0] \hbar^2}
         {8 \pi^2 s_0 m} C_3,
\label{E-relation}
\end{eqnarray}
where $f_0 = -8.42427$.
The lower limit $k > k_0$ in the $k^{-5}$ subtraction term 
avoids an ambiguity in the value of the integral over the infrared region.  
The dependence on the arbitrary wavenumber $k_0$ is cancelled by the
remaining $C_3$ term, whose coefficient depends explicitly on $k_0$.
The universal constant $f_0$ in Eq.~(\ref{E-relation}) 
was determined by matching
an expression for the energy of an Efimov trimer in the unitary limit
derived by Werner and Castin \cite{WC:1001}.

Tan's Energy Relation for fermions with two spin states
is similar to Eq.~(\ref{E-relation}) except that
there are no terms proportional to $C_3$ \cite{Tan123}.  
Combescot, Alzetto, and Leyronas proposed that such a relation 
should also apply to identical bosons \cite{CAL:0901}. 
Werner and Castin demonstrated that such a relation
does not hold for an Efimov trimer in the unitary limit \cite{WC:1001}.
Our universal relation in Eq.~(\ref{E-relation}) demonstrates 
that it fails for any state for which the 3-body contact $C_3$ is nonzero.  

Another universal relation is the Virial Theorem 
for identical bosons trapped in the harmonic potential 
$V(r) = \frac12 m \omega^2 r^2$, which was derived by Werner \cite{W:0803}:
\begin{equation}
T + U - V  = - \frac{\hbar^2}{16\pi m a} ~ C_2 - \frac{\hbar^2}{m} ~ C_3.
\label{virial}
\end{equation}
This can be derived by using dimensional analysis, 
which implies that the differential operator
$2 \omega \partial/\partial \omega  
 - a \partial/\partial a + \kappa_* \partial/\partial  \kappa_*$
is equal to 2 when acting on the total energy $E = T + U + V$. 
The Feynman-Hellmann theorem and the definitions in 
Eqs.~(\ref{E-C2,3}) imply that the three partial derivatives 
are proportional to $V$, $C_2$, and $C_3$, respectively.

One of the most important probes of ultracold atoms is
radio-frequency (rf) spectroscopy, in which an rf signal 
is used to transfer atoms to a different spin state.
In the case of fermions with two spin states, 
there are universal relations that provide sum rules 
for the rf transition rate \cite{UR:rfsum}
and control its high-frequency tail \cite{SR:0910,Braaten:2010dv}.
In the case of identical bosons, the universal relations for
rf spectroscopy also involve the 3-body contact.
The high-frequency tail of the rf transition rate is
\begin{eqnarray}
\Gamma(\omega) \longrightarrow \Omega^2 
\left[ \frac{\hbar^{1/2}}{4 \pi m^{1/2} \omega^{3/2}} ~ C_2
+ \frac{G_\textrm{rf}(\omega) \hbar}{2m \omega^2} ~ C_3 
 \right],
\label{rftail}
\end{eqnarray}
where $\Omega$ is the Rabi frequency associated with the transition.  
The transition rate is normalized so that it satisfies the sum rule
$\int \! d\omega~\Gamma (\omega) = \pi \Omega^2 N$,
where $N$ is the number of identical bosons.
The log-periodic function $G_\textrm{rf}(\omega)$ is calculated below:
\begin{equation}
G_\textrm{rf}(\omega) = B_1
+ B_2
\sin[s_0 \ln(m \omega/\hbar \kappa_*^2) + 2 \phi_{\rm rf}] ,
\label{ImF-rf}
\end{equation}
where $B_1 = 9.23$, 
$B_2 = -13.6$, and $\phi_{\rm rf} = 1.33$.

Quantum field theory is a particularly  powerful formalism for 
deriving universal relations \cite{Braaten:2008uh}.
The universal zero-range limit for identical bosons can be described 
by a quantum field theory with atom field $\psi$. 
The Hamiltonian density consists of the kinetic term for $\psi$ 
and the interaction term 
\begin{equation}
{\cal H}_\textrm{int} = 
\frac{g_2}{4 m}~d^\dagger d
+ \frac{g_3}{36 m}~t^\dagger t,
\label{H-int}
\end{equation}
where $d = \psi \psi$ and $t = \psi \psi \psi$
are local composite operators.
We set $\hbar = 1$ for simplicity.
To obtain scattering length $a$ and Efimov parameter $\kappa_*$, 
the bare coupling constants $g_2$ and $g_3$ must be tuned as functions 
of the ultraviolet momentum cutoff $\Lambda$ \cite{Bedaque:1998kg}.
If we use a sharp cutoff on the momenta of virtual particles,
the bare coupling constants must be chosen to be
\begin{subequations}
\begin{eqnarray}
g_2 &=& 8 \pi/[1/a - 2 \Lambda/\pi],
\label{g2}
\\
g_3 &=& 
- 9 g_2^2 (H + J/a \Lambda)/\Lambda^2,
\label{g3}
\\
H &=& h_0 (C - s_0 S)/(C + s_0 S),
\label{H-Lambda}
\\
J &=& [j_0 + j_1(2 S C) + j_2(C^2 - S^2)]/(C + s_0 S)^2,
\label{J-Lambda}
\end{eqnarray}
\label{g2,g3}
\end{subequations}
where $C = \cos[s_0 \ln(\Lambda/\Lambda_*)]$
and $S = \sin[s_0 \ln(\Lambda/\Lambda_*)]$.
In the renormalization prescription for $g_3$ in Eq.~(\ref{g3}), 
$H$ must be a log-periodic function of $\Lambda$.
The analytic approximation for $H$ derived in Ref.~\cite{Bedaque:1998kg}
is Eq.~(\ref{H-Lambda}) with $h_0 = 1$. 
We find that to within our numerical accuracy of about $10^{-3}$,
$H$ is given by this analytic expression multiplied by the numerical 
constant $h_0 = 0.879$.
The renormalization scale $\Lambda_*$ introduced by this prescription
differs from $\kappa_*$ by a
multiplicative factor that is only known numerically:  
$s_0 \ln (\Lambda_*/\kappa_*) = 0.971$ mod $\pi$ \cite{Braaten:2004rn}.
The function $J$ in Eq.~(\ref{g3}) is not essential for renormalization,
but it is needed to derive the Energy Relation in Eq.~(\ref{E-relation}).

Using the operational definition of $C_2$ and $C_3$ in
Eqs.~(\ref{E-C2,3}) together with the Feynman-Hellman theorem, we can
identify the 2-body and 3-body contact densities in the quantum field
theory:
\begin{subequations}
\begin{eqnarray}
{\cal C}_2 &=&  
\frac{g_2^2}{4} \left\langle d^\dagger d \right\rangle
- \frac{g_2^3}{2\Lambda^2} 
\left( H + \frac{J}{\pi} + \frac{J}{2 a \Lambda} \right)
\left\langle t^\dagger t \right\rangle,
\label{C2-density}
\\
{\cal C}_3 &=&  
-\frac{g_2^2}{8 \Lambda^2} 
\left( H' + \frac{J'}{a \Lambda} \right)
\left\langle t^\dagger t \right\rangle,
\label{C3-density}
\end{eqnarray}
\label{C2,3-density}
\end{subequations}
where $H'$ and $J'$ are the derivatives 
of $H$ and $J$ with respect to $\ln(\Lambda/\Lambda_*)$. 
The contacts $C_2$ and $C_3$ are obtained by
integrating these densities over all space.   
We have used the identity 
$(a\partial/\partial a) g_2 = g_2^2/8 \pi a$
as well as the expression for $g_3$ in Eq.~(\ref{g3}).
The condition that $C_2$ and $C_3$ have finite
limits as $\Lambda \to \infty$  implies that the matrix elements 
$\langle d^\dagger d \rangle$ and $\langle t^\dagger t \rangle$
scale as $\Lambda^2$ and $\Lambda^4$, respectively.
Thus the $\langle t^\dagger t \rangle$ term in Eq.~(\ref{C2-density})
and the $J'$ term in Eq.~(\ref{C3-density})
can be omitted unless they are multiplied by a factor of $\Lambda$.

To derive the Energy Relation in Eq.~(\ref{E-relation}),
we express the interaction energy 
$U = \int \!\! d^3R~\langle{\cal H}_\textrm{int} \rangle$ 
in terms of the contacts defined by Eqs.~(\ref{C2,3-density}):
\begin{eqnarray}
U = \frac{1/a - 2 \Lambda/\pi}{8\pi m} C_2
- \frac{2 (H + 2 J/\pi)}{m H'} C_3.
\label{U-C2,C3}
\end{eqnarray}
The coefficient of $C_2$ scales as $\Lambda$ as $\Lambda \to \infty$,
while the coefficient of $C_3$ is a log-period function of $\Lambda$.
The subtraction terms proportional to $C_2$ and $C_3$ 
in the momentum integral on the right side of Eq.~(\ref{E-relation})
can be evaluated explicitly in terms of the ultraviolet cutoff $\Lambda$.
After subtracting these terms from $U$, we find that the
dependence on the ultraviolet cutoff $\Lambda$ cancels.
The subtracted expression for $U$ reduces to the last two terms 
proportional to $C_2$ and $C_3$ on the right side of Eq.~(\ref{E-relation}).
This proves the Energy Relation and determines the constants
$A$, $\phi$, and $f_0$ in terms of the coefficients 
$h_0$, $j_0$, $j_1$, and $j_2$.

The tail of the momentum distribution in Eq.~(\ref{n-k}) 
can also be derived by using the short-distance operator product 
expansion (OPE) \cite{Braaten:2008uh}.  
The momentum distribution at wavevector $\bm{k}$
can be expressed in the quantum field theory as
\begin{equation}
n (\bm{k}) = 
\mbox{$\int$} \hbox{d}^3 R \mbox{$\int$} \hbox{d}^3r~e^{-i \bm{k} \cdot \bm{r}} 
\langle \psi^\dagger(\bm{R} \mbox{$-\frac12$} \bm{r})~
\psi(\bm{R} \mbox{$+\frac12$} \bm{r}) \rangle .
\label{n-qft}
\end{equation}
The coefficients in the OPE for
$\psi^\dagger$ and $\psi$ at equal times can be determined by matching
matrix elements between asymptotic few-body states
\cite{Braaten:2008uh}.  Alternatively they can be determined by
matching Green functions in the few-body sector. 
The simplest choice of Green functions for the
matching are those that are 
one-particle-irreducible (1PI) with respect to the atom field $\psi$
and the diatom field $d = \psi \psi$.  The resulting OPE at
large wavevector $\bm{k}$ can be expressed as
\begin{eqnarray}
&& \mbox{$\int$} \hbox{d}^3r~e^{-i \bm{k} \cdot \bm{r}}~ 
\psi^\dagger(\bm{R} \mbox{$-\frac12$} \bm{r})~
\psi(\bm{R} \mbox{$+\frac12$} \bm{r}) 
\nonumber
\\
&& = \frac{1}{k^4}~\frac{g_2^2}{4} d^\dagger d(\bm{R})
- \frac{F(k)}{k^5}~
\frac{g_2^2 H'}{8 \Lambda^2} t^\dagger t(\bm{R})
+ \ldots ,
\label{OPE:psi-psi}
\end{eqnarray}
where $F(k)$ is the log-periodic function in Eq.~(\ref{Fmom})
and the additional terms are all suppressed by at least $k^{-6}$.
The Wilson coefficient of $d^\dagger d$ was determined analytically 
by matching the diatom Green function.  The Wilson
coefficient of $t^\dagger t$ was subsequently
detemined by matching the atom+diatom Green function.  
This Green function can be expressed as a sum of loop diagrams
involving the connected atom+diatom Green function, which
can be calculated by
solving the Skorniakov-ter-Martyrosian integral equation numerically. 
Inserting the OPE in
Eq.~(\ref{OPE:psi-psi}) into the expression for the momentum
distribution $n(\bm{k})$ in Eq.~(\ref{n-qft}), 
we obtain the result for the tail of the momentum
distribution in Eq.~(\ref{n-k}).
Our direct calculation gives constants $A$ and $\phi$ 
that agree to within a few percent 
with the precise results given after Eq.~(\ref{Fmom}).  

In a quantum field theory, rf transitions of an atom to a 
different spin state can be represented 
by an operator $\psi_2^\dagger \psi$, 
where $\psi_2^\dagger$ creates an atom in the second spin state.
The rf transition rate can be expressed as 
\begin{eqnarray}
\Gamma(\omega) &=& \Omega^2 ~{\rm Im} \, i 
\mbox{$\int$} \hbox{d}t~e^{i (\omega + i \epsilon) t} 
\mbox{$\int$} \hbox{d}^3R \mbox{$\int$} \hbox{d}^3r
\nonumber\\
&& \times 
\langle \hbox{T} \psi^\dagger \psi_2(\bm{R} \mbox{$+\frac12$} \bm{r},t)~ 
 \psi_2^\dagger \psi(\bm{R} \mbox{$-\frac12$} \bm{r},0) \rangle.
\label{Gamma-psipsi}
\end{eqnarray}
We assume for simplicity
that the scattering length for the second spin state 
and the pair scattering length for the first and second spin states 
are negligible compared to $a$.  We can therefore take $\psi_2$ to be 
a noninteracting field. 
The rf transition rate at large $\omega$ can be determined by using the 
short-time operator product expansion for 
$\psi^\dagger \psi_2$ and $\psi_2^\dagger \psi$ \cite{Braaten:2010dv}.  
The leading terms 
in the OPE at large $\omega$ can be expressed as
\begin{eqnarray}
&& \mbox{$\int$} \hbox{d}t~e^{i \omega t} \mbox{$\int$} \hbox{d}^3r~
 \psi^\dagger \psi_2(\bm{R} \mbox{$+\frac12$} \bm{r},t)~ 
 \psi_2^\dagger \psi(\bm{R} \mbox{$-\frac12$} \bm{r},0)
\nonumber \\
&& = \frac{i}{\omega}~\psi^\dagger \psi(\bm{R})
+ \frac{i [(- m \omega)^{1/2} - a^{-1}]}{4 \pi m \omega^2}~
\frac{g_2^2}{4} d^\dagger d(\bm{R})
\nonumber \\
&& \hspace{2cm}
+ \frac{i F_\textrm{rf}(\omega)}{m \omega^2}~
\frac{g_2^2 H'}{8 \Lambda^2} t^\dagger t(\bm{R})
+ \ldots ,
\label{OPE:psipsi2-psi2psi}
\end{eqnarray}
where $F_\textrm{rf}(\omega)$ is a dimensionless function
and the additional terms are all suppressed by at least
$\omega^{-5/2}$.  The Wilson coefficients for
$\psi^\dagger \psi$ and $d^\dagger d$ were determined
analytically by matching the atom and diatom
Green functions, respectively.  The Wilson coefficient of 
$t^\dagger t$ was subsequently determined by
matching the atom+diatom Green function.  
The function $F_\textrm{rf}(\omega)$
in Eq.~(\ref{OPE:psipsi2-psi2psi}) has the form
\begin{eqnarray}
F_\textrm{rf}(\omega) &=& 
D_0+D_1 \ln((- m \omega)^{1/2}/\Lambda) 
\nonumber
\\
&& + D_2 \sin^2[s_0 \ln((- m \omega)^{1/2}/\kappa_*) + \phi_{\rm rf}], 
\label{Frf-omega}
\end{eqnarray}
where $\Lambda$ is the ultraviolet momentum cutoff.
Note that $F_\textrm{rf}(\omega)$ is logarithmically ultraviolet divergent. 
The numerical constants in Eq.~(\ref{Frf-omega}) are $D_0 = 0.670$,
$D_1= -2.94$, $D_2 = 1.16$, and $\phi_{\rm rf} = 1.33$.
To obtain the high-frequency tail in the rf transition rate, 
we insert the OPE in Eq.~(\ref{OPE:psipsi2-psi2psi}) into the 
expression for $\Gamma(\omega)$ in Eq.~(\ref{Gamma-psipsi})
and take the imaginary part.  
Terms that are analytic functions of $\omega$, 
such as $1/\omega$ and $1/\omega^2$,
do not contribute to the imaginary part. 
The leading terms at large $\omega$ are given by Eq.~(\ref{rftail}), 
where 
$G_\textrm{rf}(\omega) = 2~\textrm{Im} F_\textrm{rf}(\omega + i \epsilon)$.
Using our result for $F_{\textrm{rf}}(\omega)$ in Eq.~(\ref{Frf-omega}),
we obtain the result for $G_\textrm{rf}(\omega)$ in Eq.~(\ref{ImF-rf}).

There are universal relations involving the 3-body contact
for any system in which the Efimov effect arises in the 3-body problem.
Another such system consists of fermions with three spin states, 
which we label 1, 2, and 3.  In the zero-range limit, 
the interactions are described by three large pair scattering lengths 
$a_{12}$, $a_{13}$, and $a_{23}$ and an Efimov parameter $\kappa_*$.
The discrete scaling factor has the same value 
$e^{\pi/s_0} \approx 22.7$ as for identical bosons.
There are 2-body contacts $C_{12}$, $C_{13}$, and $C_{23}$
associated with each pair of spin states and a 3-body contact $C_{123}$.
The tail of the number distribution $n_1(k)$ for spin state 1
is given by Eq.~(\ref{n-k}) with $C_2$ and $C_3$ 
replaced by $C_{12}+C_{13}$ and $C_{123}$.
An operational definition of $C_{12}$ is given by Eq.~(\ref{E-C2})
with $a$ and $C_2$ replaced by $a_{12}$ and $2 C_{12}$.
An operational definition of $C_{123}$ is given by Eq.~(\ref{E-C3})
with $C_3$ replaced by $C_{123}$.  It is straightforward to derive 
the analogs of the Energy Relation in Eq.~(\ref{E-relation})
and the universal relation for the rf transition rate in Eq.~(\ref{rftail}).

Many-body systems consisting of identical bosonic atoms 
or of fermonic atoms with three or more spin states are unstable 
due to recombination into deeply-bound dimers.
The rates for these loss processes scale as $a^4$ for large $a$,
which makes it difficult to test 
universal relations for global equilibrium properties of the system,
such as the Virial Theorem in Eq.~(\ref{virial}).
However, universal relations that govern the short-time behavior of the system,
such as the tails of the momentum
distribution in Eq.~(\ref{n-k}) and the tail of the rf transition rate
in Eq.~(\ref{rftail}), can be tested experimentally by using short-time probes 
of ultracold atoms, such as those that have already been applied to 
fermions with two spin states \cite{UR:exp}.  These universal relations 
involve log-periodic functions, so they provide a new probe of
Efimov physics in ultracold atoms.

\begin{acknowledgments}
This research was supported in part by a joint grant from the 
Army Research Office and the Air Force Office of Scientific Research
and by a grant from the Department of Energy.
\end{acknowledgments}


\begin{thebibliography}{99}

\bibitem{GPS:2010}
S. Giorgini, L. Pitaevskii, and S. Stringari, 
Rev.\ Mod.\ Phys.\ {\bf 80}, 1215 (2008);
W.~Ketterle and M.W.~Zwierlein,
Making, probing and understanding ultracold Fermi gases, in
 {\it Ultracold Fermi Gases}, 
 ed. M.~Inguscio, W.~Ketterle, and C.~Salomon, 
 (IOS Press, Amsterdam) 2008.

\bibitem{Braaten:2010if}
  E.~Braaten,
  \texttt{arXiv:1008.2922}.
  
\bibitem{Tan123}
S.~Tan,
Annals of Physics {\bf 323}, 2952 (2008);
%
ibid.~{\bf 323}, 2971 (2008);
%
ibid.~{\bf 323}, 2987 (2008).

\bibitem{UR:rfsum}
M.~Punk and W.~Zwerger,
Phys.\ Rev.\ Lett.\ {\bf 99}, 170404 (2007);
%
G.~Baym, C.J.~Pethick, Z.~Yu, and M.W.~Zwierlein, 
Phys.\ Rev.\ Lett.\ {\bf 99}, 190407 (2007).

\bibitem{SR:0910}
W.~Schneider and M.~Randeria, 
Phys.\ Rev.\ A {\bf 81}, 021601(R) (2010).

\bibitem{Braaten:2010dv}
E.~Braaten, D.~Kang and L.~Platter,
Phys.\ Rev.\ Lett.\ {\bf 104}, 223004 (2010).

\bibitem{UR:photo}
F.~Werner, L.~Tarruell, and Y.~Castin, 
Eur.\ Phys.\ J.\ B {\bf 68}, 401 (2009);
%
S.~Zhang and A.J.~Leggett, 
Phys.\ Rev.\ A {\bf 79}, 023601 (2009).

\bibitem{UR:structure}
D.T.~Son and E.G.~Thompson,
Phys.\ Rev.\ A {\bf 81}, 063634 (2010);
%
H.~Hu, X.-J.~Liu, and P.D.~Drummond,
Europhys.\ Lett.\ {\bf 91}, 20005 (2010);
W.D.~Goldberger and I.Z.~Rothstein,
\texttt{arXiv:1012.5975}.



\bibitem{UR:viscosity}
E.~Taylor and M.~Randeria,
Phys.\ Rev.\ A {\bf 81}, 053610 (2010);
%
T.~Enss, R.~Haussmann, and W.~Zwerger,
Ann. Phys. (N.Y.) {\bf 326}, 770 (2011).

\bibitem{UR:exp}
E.D.~Kuhnle et al., 
Phys.\ Rev.\ Lett.\ {\bf 105}, 070402 (2010);
%
J.T.~Stewart et al.,
Phys.\ Rev.\ Lett.\ {\bf 104}, 235301 (2010).

\bibitem{Braaten:2004rn}
  E.~Braaten and H.~W.~Hammer,
  Phys.\ Rept.\  {\bf 428}, 259 (2006).

\bibitem{WC:1001}
F.~Werner and Y.~Castin, 
\texttt{arXiv:1001.0774}.
   
\bibitem{CAL:0901}
R.~Combescot, F.~Alzetto, and X.~Leyronas, 
Phys.\ Rev.\ A {\bf 79}, 053640 (2009).

\bibitem{W:0803}
F.~Werner, 
Phys.\ Rev.\ A {\bf 78}, 025601 (2008).

\bibitem{Braaten:2008uh}
  E.~Braaten and L.~Platter,
  Phys.\ Rev.\ Lett.\  {\bf 100}, 205301 (2008).
  
\bibitem{Bedaque:1998kg}
  P.~F.~Bedaque, H.~W.~Hammer and U.~van Kolck,
  Phys.\ Rev.\ Lett.\  {\bf 82}, 463 (1999).

\end{thebibliography}
\end{document}